\documentstyle[preprint,eqsecnum,oldlfont,pra,aps]{revtex}

\newcommand{\k}{{\bf{\text k}}}
\begin{document}
\draft
\title{Fluctuation effects on microphase separation\\
 in random copolymers.}
\author{ A. M. Gutin, C. D. Sfatos and E. I. Shakhnovich}
\address{Harvard University, Department of Chemistry\\
12 Oxford Street, Cambridge MA 02138}
\date{\today}
\maketitle
\begin{abstract}
We study random copolymers consisted of two kinds of monomers
with attraction between similar
kinds. The mean field analysis of this system indicates a continuous  
phase transition into a phase with periodic microdomain structure. It  
is shown that the inverse of the renormalized propagator has a  
minimum at non-zero wave-numbers. Consequently, there is an  
anomalously large contribution of fluctuations that
make the disordered phase locally stable at every finite temperature.  
However,
below a certain temperature, the ordered phase is shown to be locally  
stable and a weak first order transition is possible, similar to the  
weak crystallization theory developed by Brazovskii.
\end{abstract}

\section{Introduction}
Copolymers with attraction between similar kinds of monomers have  
received a lot of attention in the previous decade  
\cite{leibler:makromol80,bates:annu_90,bates:science_91,benoit:macrom8 
8}.
Their most important applications arise from the remarkable changes  
in their mechanical properties when they undergo a microphase  
separation transition. Regular block copolymers with well-defined  
architectures along the sequence
constitute the most important class of such materials. The study of  
these systems showed that they can undergo a temperature induced  
microphase separation transition into a variety of phases with a  
periodic domain structure and different crystalline symmetries. The  
scale of the domains is proportional to the coil size of a block  
defined by the sequence architecture and, therefore, does not depend  
on temperature.

A different class of copolymeric materials are the random -or  
statistical-
copolymers where each monomer along the sequence can be, randomly, of  
the one or the other kind \cite{gar-orl:epl597,shak:micro} or there  
is a large distribution of block lengths  
\cite{fredr:macrom92,fredr:prl91,pan-kuch:jetp91}. 

The important feature of these systems is the quenched disorder along  
the sequence. The theoretical study  on the level of mean field   
\cite{shak:micro} showed that they also undergo a temperature induced  
phase separation transition. It was shown that, near the transition  
point, the domain size depends very strongly on temperature. The  
transition from the disordered to the ordered phase has been  
predicted to be third order.

In all these cases, the finite scale of domains is a natural  
consequence
of the polymeric effect that prevents macrophase separation as would  
happen in a gas of disconnected $A$ and $B$ species with attraction  
between similar kinds. As a result of this finite scale the inverse  
of the renormalized propagator, has a minimum at finite momenta and  
the effect of fluctuations is anomalously large.

In the case of regular block copolymers, where the phase separation  
scale does not depend on temperature, the effective Hamiltonian in  
the Fourier representation is of the form \cite{leibler:makromol80}
\pagebreak
\begin{eqnarray}
{\cal H}&=&\frac{1}{2}\sum_{{\bf{\text  
k}}}[(k-k_*)^2+\tau]m({\bf{\text k}})m(-{\bf{\text k}})
\nonumber\\
&&+V^{-1}\sum_{{\bf{\text k}}_1,{\bf{\text k}}_2,{\bf{\text  
k}}_3}\lambda ({\bf{\text k}}_1,{\bf{\text k}}_2,{\bf{\text k}}_3)
m({\bf{\text k}}_1)m({\bf{\text k}}_2)m({\bf{\text  
k}}_3)m(-{\bf{\text k}}_1-{\bf{\text k}}_2-{\bf{\text k}}_3),
\end{eqnarray}
 and the order parameter is the difference between the local  
densities of each
kind of monomers.

In this Hamiltonian there is a minimum of the inverse propagator for  
${\bf{\text k}}_*\neq 0$ and the fourth order vertex has the form of  
the usual Ising
model vertex with three independent momentum vectors. The mean field  
theory for this Hamiltonian \cite{leibler:makromol80} predicts a  
continuous second order transition. In this
Hamiltonian the momentum dependence of the fourth order vertex turns  
out to be not very crucial \cite{kawasaski:macrom86} and $\lambda$  
can be taken as a
constant. Then, the Hamiltonian of the regular block copolymer can be  
mapped
onto the Hamiltonian for the Brazovskii theory of weak  
crystallization  
\cite{brazovski:jetp75,brazdzyal:jetp87,katslebmur:physrep93}.

In his seminal work \cite{brazovski:jetp75}, Brazovskii showed that  
in systems where the inverse propagator has an absolute minimum at  
${\bf{\text k}}_*\neq 0$, a continuous transition of the Landau type  
is impossible. In particular, he showed on the level of the Hartree  
approximation, that fluctuations stabilize the disordered phase and  
prevent the renormalized mass from becoming zero or negative. For  
Hamiltonians of this kind, however, he showed that, in the range  
where the Hartree approximation is valid, a phase transition is  
possible because the ordered phase becomes first locally stable and  
eventually globally stable, so a first order transition occurs.

The analogy between the Brazovskii theory and the regular  block  
copolymers was pointed out by Fredrickson and Helfand  
\cite{fredhelf:jcp87}, who calculated the corrections to the phase  
diagram of Leibler. In a later study, Dobrynin and Erukhimovich  
\cite{dobr-eruk:jphysique91} proposed a variational method to modify  
the Brazovskii theory in order to take into account the explicit  
momentum dependence of the forth order vertex reproducing the results  
of \cite{fredhelf:jcp87}.

Interesting predictions of the fluctuation effects based on the  
Brazovskii theory were also made 

by taking into account higher harmonics of the phase separation order  
parameter, regarding the behavior of some exotic structures observed  
with hexagonal cylinders
arranged in lamellar layers \cite{cruz:jcp_94}.

The explicit analogy between the Brazovskii theory and the theory of
regular block copolymers is due to the fact that the phase separation  
scale
in the latter does not depend on temperature and is fixed by the  
well-defined architecture of the sequence. In random copolymers the  
strong temperature dependence of the phase separation scale creates  
qualitative differences in the fluctuation treatment of the problem,  
as will be seen in the present work.

A first effort to describe the microphase separation transition in  
random copolymers beyond mean field was made in  
\cite{dobr-eruk:jetpl91}. It was shown that, although the inverse of  
the bare propagator has a minimum at zero momentum,
the renormalized one has a minimum at a finite momentum. In that case  
it is known from the Brazovskii theory that fluctuations make the  
disordered phase locally stable.

In all the previous mean field studies  
\cite{shak:micro,fredr:macrom92,fredr:prl91} of random copolymers  
certain terms of the Hamiltonian were ignored because they were shown  
to have a very weak effect within the limits of mean field. The  
consecutive fluctuation studies  
\cite{dobr-eruk:jphysique91,dobr-eruk:jetpl93,dobr-eruk:jetp93} also  
ignored these terms.
For this Hamiltonian it was shown recently  
\cite{sfatos:jpl94,dobr-eruk:prep} that the one-loop calculation is  
exact and higher-loop diagrams do not contribute to the Dyson  
equation in the thermodynamic limit. It was also found that, in the  
framework of the examined Hamiltonian, fluctuations destroy the  
stability of the ordered phase and the microphase separation  
transition disappears.

In the present work we examine the complete effective Hamiltonian for  
random copolymers with strong short range correlations of monomer  
kinds along the sequence, {\em i.e.}, with a wide distribution of  
block lengths.
We show that the previously omitted terms play an important role in  
restoring the phase transition. 

Due to the presence of these extra terms the ordered phase becomes  
stable
at some temperature and the system can undergo a weak first order  
transition
to a phase with  periodic microdomain structure. The period of the  
domains
and the amplitude of the separation below the transition temperature  
are the same as those predicted by mean field.

\section{The Model and Mean field theory}
The random two-letter copolymer is described by a microscopic  
Hamiltonian that takes into account the self-interactions between  
monomers as
\begin{equation}
{\cal H}=\frac{1}{2}\sum_{i\neq j}
B_{ij}U({\bf {\text r}}_{i}-{\bf {\text r}}_{j})
\end{equation}
where the conformation of the polymer is described by the coordinates  
of 

its monomers $\{{\bf {\text r}}_{i}\}$ and $U({\bf {\text r}})$ is a  
short-range
potential. The binary interaction virial coefficient is given  
\cite{obuk:jpa86} by 

\begin{equation}
B_{ij}=
\chi\sigma_{i}\sigma_{j}
+c_1(\sigma_{i}+\sigma_{j})+c_2
\end{equation}

The sequence of monomers is described by a quenched set of
random values
$\{\sigma_{i}\}$ with equal probabilities for the two types of  
monomers; $\sigma_{i}=1$ if monomer $i$ is of type $A$ and  
$\sigma_{i}=-1$ if it is of
type $B$\@. When the interactions between similar monomers are equal  
$(B_{AA}=B_{BB})$ then $c_1=0$. The composite Flory parameter  
\mbox{$\chi=(B_{AA}+B_{BB})/2-B_{AB}$} will be negative in the case  
of interest where similar monomers attract each other. The constant  
$c_2$ corresponds to an overall attraction in the two body term. This  
overall attraction, in combination with a three-body repulsion term  
which is not explicitly introduced in the Hamiltonian, are known to  
lead the polymer to a compact state  with constant density  
\cite{grosberg:book}. This well-studied effect is purely  
homopolymeric and will not be considered in the present study where  
only heteropolymeric effects are taken into account.

In this model we consider that a monomer of one kind is followed,  
with high probability, by a monomer of the same kind.  
Correspondingly, the correlations between kinds of monomers decay as
\begin{equation}
\langle\sigma_i\sigma_j\rangle
\sim e^{-|i-j|/l},
\end{equation}
where $l\gg 1$ represents the average block length with monomers of  
one kind.

The polymeric effect is explicitly introduced through the elastic  
term \cite{grosberg:book} that constrains the position of consecutive  
monomers at an average distance 1:
\begin{equation}
g({\bf {\text r}}_{i+1}- {\bf {\text r}}_{i} )=
\frac{1}{(2\pi)^{3/2}}
\exp\biggl[ -\frac{({\bf {\text r}}_{i+1}-{\bf {\text r}}_{i})^{2}}
{2} \biggl]
\end{equation}

In this system, attraction between similar types of monomers  
generates an energetic preference for phase separation. In the  
absence of polymeric bonds,
a complete phase separation is taking place. In the presence of  
polymeric bonds, though, we can only expect a microphase separation.  
This phase separation is described by the order parameter  
\cite{gar-orl:epl597}
\begin{equation}
m({\bf{\text R}})=\sum_i\sigma_i\delta({\bf{\text r}}_i-{\bf{\text  
R}})
\end{equation}
which corresponds to the difference $\rho_A-\rho_B$ between the  
densities  of the two types of monomers. In the present study we  
refer to the case where there is an equal amount of $A$ and $B$ types  
which is usually referred to as composition $f=1/2$.

The effective Hamiltonian  is of the form  
\cite{shak:micro,fredr:macrom92}
\begin{eqnarray}
{\cal H}=\frac{1}{2}\sum_{{\bf{\text k}}\neq 0}({\bf{\text  
k}}^2+\tau)m({\bf{\text k}})m(-{\bf{\text k}})
&&+V^{-1}\lambda_1\sum_{{\bf{\text k}}_1,{\bf{\text k}}_2\neq 0}
\frac{m({\bf{\text k}}_1)m(-{\bf{\text k}}_1)m({\bf{\text  
k}}_2)m(-{\bf{\text k}}_2)}
{{\bf{\text k}}_1^2+{\bf{\text k}}_2^2}
\nonumber\\
&&+V^{-1}\lambda_2 \sum_{{\bf{\text k}}_1,{\bf{\text k}}_2,{\bf{\text  
k}}_3}
m({\bf{\text k}}_1)m({\bf{\text k}}_2)m({\bf{\text  
k}}_3)m(-{\bf{\text k}}_1-{\bf{\text k}}_2-{\bf{\text k}}_3)
\end{eqnarray}
where  $V$ is the volume of the system and the Fourier transform is  
defined as 

$m({\bf{\text k}})=(1/\sqrt{V})\int d{\bf{\text R}} m({\bf{\text  
R}})e^{i{\bf{\text k}}\cdot{\bf{\text R}}}$.

The  coefficient $\lambda_1$ of the first vertex is inversely  
proportional to the square of the average block length $l$, {\em  
i.e.}, $\lambda_1\sim 1/l^2$. 

This vertex corresponds to the loss of entropy due to polymeric bonds  
that connect
different blocks with each other. We can see immediately that, due to  
this term, long wave-length modes become unfavorable. On the other  
hand, very short wave-length modes are unfavorable due to the   
surface  tension contribution in the second order term.

The $\lambda_2$ vertex has the usual form of the vertex in the Ising  
model effective Hamiltonian. This term
is due to the discrete values $\pm 1$ of the  sequence  
$\{\sigma_i\}$. In an annealed system, as for example in a binary  
alloy, where the quenched polymeric bonds are absent, this is the  
entropic fourth order term that provides stability of the ordered  
phase with macroscopic phase separation into two domains. In the case  
of random copolymers the coefficient
of this vertex is $\lambda_2\sim 1/l$, {\em i.e.}, it is of order  
unity per
typical block in units of $k_BT$. The momentum dependence of this  
term can be neglected for scales larger than the coil size of the  
average block
which is of order $l^{1/2}$. This is the term omitted from all  
previous mean field studies. It is seen immediately that this term  
becomes comparable with the 

other vertex $\lambda_1/k^2$ only at scales $k\sim 1/l^{1/2}$ and  
therefore
it was argued that it can be omitted for larger domain scales  
predicted by mean field.
 Then, it was shown \cite{shak:micro} that the mean field solution  
can be taken in the form

\begin{equation}
m({\bf{\text k}})=m_0\sqrt{V/2}
\Bigl( \Delta ({\bf{\text k}}-{\bf{\text k}}_0)+\Delta ({\bf{\text  
k}}+{\bf{\text k}}_0)
\Bigl),
\end{equation}
where $\Delta$ is Kronecker's delta. This solution corresponds to the  
lamellar phase. The mean field amplitude $m_0$ and the frequency  
$k_0$ can be determined by minimization of (2.6) to be

\begin{equation}
m_0=0\ \ \ \ \text{for}\ \ \tau >0;\ \ \ \ \ 

m_0=-\frac{\tau}{3\sqrt{\lambda_1}},
\ \ \ \ {\bf{\text k}}_0^2=-\frac{\tau}{3},
\ \ \ \ \text{for}\ \ \tau <0,
\end{equation}
where the effect from $\lambda_2$ is negligible because $k_0$ is very  
small near the transition.

Thus, the mean field theory predicts a continuous  phase transition  
at $\tau=0$. This transition was found to be third order. The  
effective Hamiltonian which contains only the heteropolymeric
vertex is equivalent to a Gaussian sequence model, {\em i.e.}, a  
quenched sequence of Gaussian variables $\{\sigma_i\}$ instead of  
discrete ones $\pm 1$. This approximation does not change the mean  
field results but
is crucial for the fluctuation analysis that will be performed in the  
following sections. We will show in Sec.\ IV that this is the only  
term that provides the stability of the order phase. 

We will refer to the effective Hamiltonian without the $\lambda_2$  
vertex as the Gaussian sequence approximation.

\section{The Gaussian  Sequence Approximation.}

In this section we will consider the effective Hamiltonian in the 

Gaussian sequence approximation, describing the microphase separation  
transition in random copolymers, as
\begin{equation}
{\cal H}=\frac{1}{2}
\sum_{{\bf{\text k}}\neq 0}({\bf{\text k}}^2+\tau)
m({\bf{\text k}})m(-{\bf{\text k}})
+V^{-1}\lambda_1\sum_{{\bf{\text k}}_1,{\bf{\text k}}_2\neq 0}
\frac{m({\bf{\text k}}_1)m(-{\bf{\text k}}_1)
m({\bf{\text k}}_2)m(-{\bf{\text k}}_2)}
{{\bf{\text k}}_1^2+{\bf{\text k}}_2^2}
\end{equation}

This is the Hamiltonian that has been used in all previous mean field  
studies.
 Shortly after the derivation of this Hamiltonian
it was observed \cite{dobr-eruk:jetpl91} that a renormalization of  
the Green
function on the one-loop level changes its form qualitatively. The  
one-loop Dyson equation for the  Green function is
\begin{equation}
G^{-1}({\bf{\text k}})=( k^2+\tau)+V^{-1}\sum_{{\bf{\text  
k}}_1}\frac{4\lambda_1}{k_1^2+k^2}G({\bf{\text k}}_1).
\end{equation}
By substituting even the bare Green function into the integral of  
(3.2) for $\tau >0$, the renormalized Green function in three  
dimensions is
\begin{equation}
G^{-1}({\bf{\text k}})=k^2+\tau 

+\frac{2\lambda_1/\pi}{\tau^{1/2}+k}, 

\end{equation}
with $k=|{\bf{\text k}}|$. We see immediately that there is a minimum  
of $G^{-1}({\bf{\text k}})$ at some ${\bf{\text k}}_*\neq 0$. It was,  
therefore, proposed that the form of the renormalized Green function  
can be described by the form used in the weak crystallization theory  
\cite{brazovski:jetp75}
\begin{equation}
G^{-1}({\bf{\text k}})=C(k-k_*)^2+r.
\end{equation}
This approximation is good for $r\ll {\bf{\text k}}_*^2$.
By substitution of this into (5) it is found that
\begin{equation}
r=\tau +\frac{3\lambda_1}{2\pi (Cr)^{1/2}},\ \ \    
k_*^2=\frac{\lambda_1}{2\pi(Cr)^{1/2}},\ \ \ 

C=2.
\end{equation}
According to the first relation in (3.5), the renormalized mass $r$,  
cannot become zero except for $\tau=-\infty$ which corresponds to  
$T=0$.  Therefore
the disordered phase never looses stability as in the Brazovskii  
theory. This               result is quite general and is due to the  
fact that the integral corresponding to the one loop correction,
\begin{equation}
\int\frac{d^3{\bf{\text k}}_1}{(k^2+k_1^2)[(k_1-k_*)^2+r]},
\end{equation}
is divergent as $r\rightarrow 0$. On the basis of this evidence it  
was assumed  that this system will have a first order transition of  
the Brazovskii type. A more careful study, however, was done in  
\cite{sfatos:jpl94,dobr-eruk:prep} and is presented briefly in the  
rest of this section.

Consider the diagrams that contribute to a perturbation expansion.
The contribution of the one-loop diagram shown in Fig.\ 1(a) to the   
Dyson equation is of order

\begin{equation}
V^{-1}\lambda_1 \sum_{{\bf{\text k}}_1}
\frac{G({\bf{\text k}}_1)}
{k_1^2+k^2}
=V^{-1}\lambda_1\frac{V}{(2\pi)^3}
\int d^3{\bf{\text k}}_1\frac{G({\bf{\text k}}_1)}
{k_1^2+k^2}\sim 1.
\end{equation}

The two-loop diagram shown in Fig.\ 1(b) is of order
\begin{equation}
\lambda_1^2 V^{-2} G({\bf{\text k}})\frac{V}{(2\pi)^3}\int  
d^3{\bf{\text k}}_1 

\frac{G^2({\bf{\text k}}_1)}{(k_1^2+k^2)^2}\sim V^{-1},
\end{equation}
and should be neglected in the thermodynamic limit. We can easily see  
that all higher-loop diagrams do not contribute for the same reasons.

From the above remarks we conclude that the one-loop Dyson equation  
given in (3.2) is exact and the lack of continuous transition for the  
effective Hamiltonian (3.1) is a general result that does not depend  
on the smallness of the parameter $\lambda_1$.

On the premises discussed above, we calculate here the stability of  
the ordered phase. If we assume that the symmetry is broken as  
described by
(2.7) we need to write down the free energy functional ${\cal  
H}\{m_0,{\bf{\text k}}_0;\psi\}$. The order parameter is
\begin{equation}
m({\bf{\text k}})=\Biggl\{
\begin{array}{lll}
m_0\sqrt{V/2}&\text{for}
&{\bf{\text k}}=\pm{\bf{\text k}}_0\\
\psi({\bf{\text k}})&\text{for}
&{\bf{\text k}}\neq\pm{\bf{\text k}}_0.
\end{array}
\end{equation}

Fluctuations of the mode ${\bf{\text k}}={\bf{\text k}}_0$  are of  
order 1 and are ignored compared to the mean field $m({\bf{\text  
k}})\sim \sqrt{V}$. Fluctuations of other modes are denoted by  
$\psi$. Then the free energy functional becomes
\begin{eqnarray}
{\cal H}'\{m_0,{\bf{\text k}}_0;\psi\}
={\cal H}\{m_0,{\bf{\text k}}_0\}
&+&\frac{1}{2}\sum_{{\bf{\text k}}\neq {\bf{\text k}}_0}
\biggl(k^2+\tau +\frac{4\lambda_1 m_0^2}
{k_0^2+k^2}
\biggl)\psi({\bf{\text k}})\psi(-{\bf{\text k}})
\nonumber\\
&+&\lambda_1 

V^{-1}\sum_{{\bf{\text k}}_1,{\bf{\text k}}_2\neq {\bf{\text k}}_0}
\frac{\psi({\bf{\text k}}_1)\psi(-{\bf{\text k}}_1)
\psi({\bf{\text k}}_2)\psi(-{\bf{\text k}}_2)}
{k_1^2+k_2^2},
\end{eqnarray}
where ${\cal H}\{m_0,{\bf{\text k}}_0\}$ is the value of the  
Hamiltonian (3.1) if we substitute the mean field solution in the  
form given by (2.7).

The classical field values $m_0, {\bf{\text k}}_0$ are found from the  
equation of state given by the thermodynamic  relations
\begin{equation}
\left\langle
\frac{\partial {\cal H}'\{m_0,{\bf{\text k}}_0;\psi\}}{\partial m_0}
\right\rangle =0
\ \ \ \text{and}\ \ \ 

\left\langle
\frac{\partial {\cal H}'\{m_0,{\bf{\text k}}_0;\psi\}}{\partial  
{\bf{\text k}}_0}
\right\rangle =0.
\end{equation}
The first equation reads
\begin{equation}
(k_0^2+\tau )m_0+\frac{2\lambda}{k_0^2}m_0^3
+4\lambda m_0V^{-1}\sum_{{\bf{\text k}}\neq{\bf{\text k}}_0}
\frac{\langle\psi({\bf{\text k}})\psi(-{\bf{\text k}})\rangle}
{k_0^2+k^2}=0
\end{equation}
where $\langle\psi({\bf{\text k}})\psi(-{\bf{\text  
k}})\rangle=G({\bf{\text k}})$. The Dyson equation for the ordered  
phase for ${\bf{\text k}}\neq{\bf{\text k}}_0$ is
\begin{equation}
G^{-1}({\bf{\text k}})=k^2+\tau 

+\frac{4\lambda m_0^2}{k_0^2+k^2}
+4\lambda V^{-1}
\sum_{{\bf{\text k}}_1\neq{\bf{\text k}}_0}
\frac{G({\bf{\text k}}_1)}{k_1^2+k^2}.
\end{equation}
This equation is exact since higher-loop diagrams are subdominant in  
$V$  due to the peculiar symmetry of our vertex.

Then, by comparison of Eqs.\ (3.12) and (3.13) we see that  
$G({\bf{\text k}})$ depends only
on the modulus $k$ and
\begin{equation}
G^{-1}(k_0)m_0=0.
\end{equation}
The minimum value of $G^{-1}({\bf{\text k}})$ is positive since if  
$G^{-1}=0$ the integral of (16) diverges. Then
$G^{-1}(k_0)>0$ and therefore, according to Eq.\ (3.14), $m_0=0$.
We see that we cannot have a stable solution with $m_0\neq 0$ and $  
k_0\neq 0$ for the Hamiltonian (3.1).


\section{The complete Hamiltonian}

\subsection{The Ginsburg Criteria}

In the previous section we saw that, although on the level of mean  
field the results should not be affected by the omission of the  
second vertex $\lambda_2$, the fluctuation analysis using only the  
heteropolymeric vertex $\lambda_1$ results in the absence of any kind  
of  microphase separation transition. This result is expectable even  
from mean field estimates
if we notice the following. The inverse propagator for the ordered  
phase calculated
on the level of mean field in the absence of $\lambda_2$ is
\begin{equation}
G^{-1}({\bf{\text k}})=
k^2+\tau +\frac{4\lambda_1 m_0^2}
{k_0^2+k^2}.
\end{equation}
By substitution of the mean field values (2.8) we see that the  
inverse propagator has a minimum at
${\bf{\text k}}_0$.
Then the bare mass in the ordered phase is given by
\begin{equation}
r=\tau +k_0^2+\frac{2\lambda_1}{k_0^2}m_0^2.
\end{equation}
By substitution of the mean field values (2.8) for $m_0$ and $k_0$  
this gives $r=0$.
The consideration of the complete Hamiltonian (2.6) is then  
necessary. If $\lambda_2$ is included we see that the bare mass  
becomes positive and the ordered
phase is stable independently of the relative smallness of  
$\lambda_2$ compared
to $\lambda_1/k_0^2$. It is straightforward to show that the mass for  
the ordered phase on the level of mean field is

\begin{equation}
r=8\lambda_2 m_0^2\ \ \ 

\text{or}\ \ \ r\sim \tau^2l.
\end{equation}

We now shift our attention to the effect of fluctuations. The point  
at which fluctuations
become important can be estimated by comparing the contribution of  
the diagram of Fig.\ 2(a) to the $\Gamma^{(4)}$ function  with the  
bare vertex value. The
vertex $\lambda_1/k_0^2$ gives larger contribution than the  
equivalent of $\lambda_2$ at a given temperature and therefore the
Ginsburg criterion for the validity of mean field should be  
calculated
by the diagram of  Fig.\ 2(a) for the $\lambda_1$ vertex.

 We first consider the disordered phase. In this case the Ginsburg  
criterion
is given by
\begin{equation}
\lambda_1^2\int\frac{k^2 dk}
{(k^2+k_1^2)(k^2+k_2^2)(k^2+\tau)^2}
\ll\frac{\lambda_1}{k_1^2+k_2^2}.
\end{equation}
This is satisfied for any $k_1, k_2$ if
\begin{equation}
\tau\gg \frac{1}{l^{4/3}}
\end{equation}

For the ordered phase the Ginsburg criterion is equivalent to  
$\lambda_1/r^{3/2}\ll 1$ which due to Eq.\ (4.3) becomes

\begin{equation}
-\tau\gg \frac{1}{l^{7/6}}.
\end{equation}
From the other hand we can see that the mean field value for $m_0$  
becomes
of order one at $\tau\sim 1/l$ and the Landau expansion breaks down.
Therefore the region where mean field is valid for the ordered phase  
is
\begin{equation}
\frac{1}{l^{7/6}}\ll -\tau\ll\frac{1}{l}.
\end{equation}
Since mean field is correct in this region we can conclude that {\em  
there is} a phase transition between the disordered phase with  
$m_0=0$ and the ordered
phase with $m_0\neq 0$.

Analogously, fluctuation corrections arising from the $\lambda_2$  
vertex
can be neglected for $\lambda_2 k_0^2/r^{3/2}\ll 1$. This gives the  
Ginsburg
criterion for this vertex in the ordered phase
\begin{equation}
-\tau\gg\frac{1}{l^{5/4}},
\end{equation}
which is smaller than the Ginsburg value of $-\tau$
for $\lambda_1$ as anticipated above.

\subsection{The Effect of Fluctuations}

According to the derived criteria there is a region of temperatures
\begin{equation}
\frac{1}{l^{5/4}}\ll -\tau\stackrel{<}{_\sim}\frac{1}{l^{7/6}}
\end{equation}
where the Dyson equation in the ordered phase should be described by
\begin{equation}
G^{-1}({\bf{\text k}})=k^2+\tau
+\frac{4\lambda_1 m_0^2}
{k_0^2+k^2}
+12\lambda_2m_0^2
+4\lambda_1 V^{-1}
\sum_{{\bf{\text k}}_1\neq{\bf{\text k}}_0}
\frac{G({\bf{\text k}}_1)}
{k_1^2+k^2},
\end{equation}
because the one-loop contribution of the vertex $\lambda_1$ is exact  
according to the analysis given in Sec.\ III. The  equation obtained  
from the variation of the free energy expression with respect to the  
amplitude $m_0$ is
\begin{equation}
({\bf{\text k}}_0^2+\tau )m_0
+\frac{2\lambda_1}{{\bf{\text k}}_0^2}m_0^3
+4\lambda_2 m_0^3+4\lambda_1 m_0V^{-1}
\sum_{{\bf{\text k}}\neq{\bf{\text k}}_0}
\frac{G({\bf{\text k}})}
{{\bf{\text k}}_0^2+{\bf{\text k}}^2}=0.
\end{equation}
By comparison of Eqs.\ (4.10) and (4.11) we see that there is a  
solution for
the field amplitude
\begin{equation}
m_0^2=\frac{G^{-1}({\bf{\text k}}_0)}{8\lambda_2}.
\end{equation}

From the minimization of the free energy with respect to the  
parameter
$k_0^2$ we get
\begin{equation}
\frac{m_0^2}{2}\biggl[
1-\frac{\lambda_1 m_0^2}{k_0^4}
-4\lambda_1 V^{-1}
\sum_{{\bf{\text k}}\neq{\bf{\text k}}_0}
\frac{G(\k)}
{(k_0^2+k^2)^2}\biggl]=0.
\end{equation} 

This equation coincides with the equation for the minimum $k_*$ of  
the inverse propagator by differentiation of the Dyson equation  
(4.10) for the ordered phase and
$k_*$ coincides with $k_0$.
By taking $G^{-1}({\bf{\text k}})=(k-k_0)^2+r$ and using Brazovskii's  
estimate for the integrals it follows from Eq.\ (4.10)
\begin{equation}
r=k_0^2+\tau+\frac{2\lambda_1}{k_0^2}m_0^2
+\frac{3\lambda_1}{2\pi r^{1/2}}.
\end{equation}
Substituting Eq.\ (4.12) in  Eq.\ (4.14) we obtain
\begin{equation}
-(\tau +{\bf{\text k}}_0^2)=
\biggl(\frac{\lambda_1}{4k_0^2\lambda_2}-1
\biggl) r+\frac{3\lambda_1}{2\pi r^{1/2}}.
\end{equation}
Since we are considering scales much larger than
the coil size of the average  block we always have  
$\lambda_1/(k_0^2\lambda_2)\gg 1$
and the Dyson equation for the ordered phase has the form
\begin{equation}
-(\tau +{\bf{\text k}}_0^2)=
\frac{\lambda_1}{4k_0^2\lambda_2}r
+\frac{3\lambda_1}{2\pi r^{1/2}}.
\end{equation}
We notice that the second term of the right hand side of (4.16)
is small compared to $-\tau$ for $-\tau\gg 1/l^{5/4}$.  The same
condition is also valid in order to neglect the contribution of
the last term in (4.13). Consequently,
the mean field results for $r, m_0$ and $k_0$ will hold true in 

the region (4.9).

As we approach $-\tau\sim 1/l^{5/4}$ the three terms of
(4.16) become of the same order. 

It easy to show that for $-\tau >\tau_c$
and, correspondingly, for $r>r_c$ this equation has real solutions  
for $r$
where
\begin{equation}
r_c\sim\frac{1}{l^{3/2}}\ \ \ \ \text{and}
\ \ \ \ \ \tau_c\sim\frac{1}{l^{5/4}},
\end{equation}
and the solution with $r$ growing as $-\tau$ grows corresponds
to the minimum of the thermodynamic potential. Correspondingly, for
$-\tau <\tau_c$ Eq.\ (4.16) does not  have a solution. This means  
that
in that region the ordered phase is locally unstable.

 We can easily see from the first equation in (3.5) that
the disordered phase does not loose local
stability at $-\tau\sim 1/l^{5/4}$. Then  there must be a first order
phase transition as in the theory of weak crystallization of  
Brazovskii.
It can be shown rigorously with Brazovskii's method 

\cite{brazovski:jetp75} that at the temperature
of the order predicted by (4.17) the ordered phase becomes
globally stable.
The jump in the amplitude, according to (4.12), is given by
\begin{equation}
m_0\sim\frac{1}{l^{1/4}}\ \ \ \ \ \text{and}
\ \ \ \ \ k_0^2\sim\frac{1}{l^{5/4}}.
\end{equation}

We note that the order of the transition temperature coincides with  
the Ginsburg temperature
for the vertex $\lambda_2$ predicted by Eq.\ (4.8). However, one-loop  
corrections to the Dyson equation due to this vertex need not be  
considered
because the divergent contribution of the $\lambda_1$ vertex is much  
larger.
It is also easy to show that at $-\tau\sim 1/l^{5/4}$ the higher-loop
terms for $\lambda_2$ do not contribute.

As explained in Sec.\ III, the one-loop correction for the vertex  
$\lambda_1$
is not an approximation but the exact contribution of this vertex to  
the self-energy. We must also mention that for the renormalization of  
the four-point function $\Gamma_1^{(4)}$ for this vertex, only the  
channel of
Fig.\ 2(a) contributes to the geometric progression for the ladder  
diagrams
since the channel of Fig.\ 2(b) must be neglected according to the  
arguments of
Sec.\ III. In contrast, for the vertex $\lambda_2$ Brazovskii showed  
that there are two channels contributing to the renormalization of  
$\Gamma_2^{(4)}$ as shown in Fig.\ 2(c)-(d) and
the geometric progression for the renormalized four-point function  
has a special form given by
\begin{equation}
\Gamma_2^{(4)}=\frac{1-\lambda_2\Pi}{1+\lambda_2\Pi},
\end{equation}
where $\Pi\sim k_0^2/r^{3/2}$. It is seen that when  
$\lambda_2k_0^2/r^{3/2}>1$
$\Gamma_2^{(4)}$ changes sign which does not  happen for the  
four-point function $\Gamma_1^{(4)}$. This change of sign can be  
interpreted
as a hint for the existence of an inflection in the thermodynamic  
potential
which should exist at the temperature where the ordered phase becomes  
locally stable and below.


\section{Discussion}

We investigated the effect of thermal fluctuations on the microphase  
separation transition in a melt of random copolymers. We considered a  
quenched sequence consisted of an equal
amount  of monomers of two kinds in the thermodynamic limit.  

The model assumes that neighboring monomers are with high probability
of the same kind
with average block length $l\gg 1$ and a wide distribution of block  
lengths.
Our analysis indicates that there is a weak first order phase  
transition between a disordered and an ordered phase with periodic  
microdomain structure instead of a continuous third order one  
predicted by mean field.

The effective Hamiltonian (2.6) for this system contains two  
fourth-order vertices with completely different forms. The vertex  
labeled $\lambda_1$ is solely due to the polymeric effect and  
prohibits macroscopic phase separation. The vertex $\lambda_2$ which  
has the usual Ising model form is due to the discrete values $\pm 1$  
for the sequence labels $\{\sigma_i\}$. In mean field, the  
contribution of the vertex  $\lambda_2$ is negligible around the  
transition point and can be  omitted. An exact analysis of  
fluctuations in the absence of  vertex $\lambda_2$ is possible and  
shows that the disordered phase is always locally stable. At the same  
time the
ordered phase is locally unstable and the phase transition is  
impossible. 

Then the effect of the vertex $\lambda_2$ has to be taken into  
account.

For the complete Hamiltonian (2.6), we showed that the mean field  
predictions
\begin{equation}
m_0\sim -\tau l \ \ \ \
\text{and}\ \ \ \ k_0^2\sim -\tau,
\ \ \ \  \text{for} \ \ \tau <0,
\end{equation}
are correct in the region
\begin{equation}
\frac{1}{l^{5/4}}\stackrel{<}{_\sim} -\tau\ll\frac{1}{l}.
\end{equation}
The weak first order transition takes place at $-\tau\sim 1/l^{5/4}$
with a jump in the amplitude
\begin{equation}
\Delta m_0\sim \frac{1}{l^{1/4}}
\ \ \ \ \text{and}
\ \ \ \ k_0\sim\frac{1}{l^{5/8}}.
\end{equation}

The longer the average block length $l$, the higher the 

temperature of transition and the smaller the jump in the  phase  
separation
amplitude at the point of the weak first order transition.

In the case of regular block copolymers, the jump in the  
corresponding
fluctuationally induced first order transition predicted in  
\cite{fredhelf:jcp87} is $\Delta m_0\sim 1/l_b^{1/6}$, where $l_b$  
the fixed block size of the sequence. We see that the jump predicted  
for random copolymers is smaller. The temperature of transition
in regular block copolymers is predicted at $-\tau\sim 1/l_b^{4/3}$.  
We see that
the corresponding shift for random copolymers is also smaller.
We notice that in random copolymers the domain size near transition
$k_0^{-1}\sim l^{5/8}$
is much larger than the corresponding domain size in regular block  
copolymers
$k_b^{-1}\sim l_b^{1/2}$. The amplitude of the transition at the  
ordered phase in regular block copolymers $m_0\sim (-\tau l_b)^{1/2}$  
is 

larger than the predicted amplitude of the random copolymer  
microphase separation at the same temperature.

We also see that, when there are no correlations along the sequence  
{\em i.e.} $l\sim1$,
the predicted transition occurs at a region where the Landau
expansion fails. The physics of this
system is also described by a freezing transition into a phase where  
a small number of conformations is thermodynamically  dominant  
\cite{shak:natlett} and depends strongly on the flexibility of the  
chain \cite{sfatos:pre93}.

It is also worth noticing that in the absence of the vertex  
$\lambda_2$,
the vertex $\lambda_1$ does not have any preference for the different
Bravais lattices \cite{dobr-eruk:jetpl91}. This degeneracy
of the polymeric vertex $\lambda_1$ is related to the instability
of the ordered phase predicted on the basis of only this vertex.
All the structure dependence comes from the vertex $\lambda_2$.  
Therefore the analysis of structures other than the lamellar
is expected to be similar to that of Brazovskii. In the case of  
composition $f=1/2$, there is no cubic term that breaks the inversion  
symmetry. Although other structures are locally stable the lamellar  
would retain global stability. For compositions $f\neq 1/2$ other  
structures like BCC and hexagonal cylinders may become globally  
stable and transitions between them may be possible.

In all the calculations the integral for the one-loop correction
of Fig.\ 1(a) has been estimated \cite{brazovski:jetp75}
to be of order $\lambda k_0^2/r^{1/2}$.
In fact this estimate is correct at $r\ll k_0^4$. For $k_0^4\ll r\ll  
k_0^2$   the estimate
\begin{equation}
\lambda\int\frac{k^2dk}{(k-k_0)^2+r}
\sim\lambda D+\frac{\lambda k_0^2}{r^{1/2}}
\end{equation}
describes completely the behavior of the integral and its
derivatives, where $D$ is a numerical constant that depends on the  
cut-off as in the Ising case. The $\lambda D$ term simply shifts
$\tau$.

\acknowledgements
We would like to thank A.V. Dobrynin and I.Ya.\ Erukhimovich for  
stimulating
discussions and for sending to us the preprint of Ref.\  
\cite{dobr-eruk:prep}. This work was supported by the David and  
Lucille Packard Fund.

\begin{figure}
\caption{One-loop and two-loop Feynman diagram contribution to the  
self-energy.}
\end{figure}
\begin{figure}
\caption{(a)-(b) One-loop diagram contribution to the four-point  
function $\Gamma_1^{(4)}$ for vertex $\lambda_1$. (c)-(d) One-loop  
diagram contribution to the four-point function $\Gamma_2^{(4)}$ for  
vertex $\lambda_2$.}
\end{figure}
\end{document}